# Impact of Surface and Pore Characteristics on Fatigue Life of Laser Powder Bed Fusion Ti-6Al-4V Alloy Described by Neural Network Models

Seunghyun Moon[1], Ruimin Ma[1], Ross Attardo[2], Charles Tomonto[2], Mark Nordin[3], Paul Wheelock[3], Michael Glavicic[3], Maxwell Layman[3], Richard Billo[4], Tengfei Luo[1,5,*]

1. Department of Aerospace and Mechanical Engineering, University of Notre Dame, IN 46556

2. 3D Printing Center, Johnson & Johnson, Miami, FL 33126

3. Rolls-Royce Corporation, 450 S. Meridian St., Indianapolis, IN 46225

4. Department of Computer Science and Engineering, University of Notre Dame, IN 46556

5. Department of Chemical and Biomolecular Engineering, University of Notre Dame, IN 46556

* corresponding author: tluo@nd.edu

**Abstract:**

In this study, the effects of surface roughness and pore characteristics on fatigue lives of laser powder bed fusion (LPBF) Ti-6Al-4V parts were investigated. The 197 fatigue bars were printed using the same laser power but with varied scanning speeds. These actions led to variations in the geometries of microscale pores, and such variations were characterized using micro-computed tomography. To generate differences in surface roughness in fatigue bars, half of the samples were grit-blasted and the other half machined. Fatigue behaviors were analyzed with respect to surface roughness and statistics of the pores. For the grit-blasted samples, the contour laser scan in the LPBF strategy led to a pore-depletion zone isolating surface and internal pores with different features. For the machined samples, where surface pores resemble internal pores, the fatigue life was highly correlated with the average pore size and projected pore area in the plane perpendicular to the stress direction. Finally, a machine learning model using a drop-out neural network (DONN) was employed to establish a link between surface and pore features to the fatigue data (*logN*), and good prediction accuracy was demonstrated. Besides predicting fatigue lives, the DONN can also estimate the prediction uncertainty.

# 1. Introduction

Laser powder bed fusion (LPBF) is an additive manufacturing (AM) technique that uses high-power laser heating to melt and fuse fine metallic particles layer-by-layer in a powder bed[1-3]. It offers an excellent opportunity for providing more versatility and efficiency in diverse environments as sustainment facilities and hospitals for asset replacement[4,5]. LPBF can result in drastic improvements in shortening component development/refurbishment time, reducing the length and complexity of the supply chain, reducing inventory levels, replacing a medical device or returning an asset to operational service in a much shorter time, and ultimately resulting in an overall reduction of the cost of asset replacement and sustainment, and improved service.

However, uncertainty and perceived risk associated with the mechanical properties of the products from this new technology have slowed its adoption to take advantage of its merits. Uncertainties are typically associated with concerns surrounding the processing and resultant material of a component. Factors such as laser power, scanning speed, powder contamination, process interruptions, unremoved powder, laser instability and surface finishing can all impact the microstructure of the printed material, leading to engineering uncertainty in the mechanical performance of the printed part[6-10]. As a result, industry takes a relatively conservative stance in the acceptance of LPBF materials and printed components, and thus slow the widespread adoption of this advanced manufacturing technology for innovations. A number of studies have been conducted to overcome the engineering uncertainty or to understand the underlying failure mechanism for LPBF parts. For instance, Ren *et al.*[11] have reported that AM Ti-6Al-4V (Ti64) parts produced by high power laser can improve the fatigue lives after solution treatment and aging. Chan *et al.*[12] and Greitemeier *et al.*[8] focused on the aspects related to the influence of surface roughness on the fatigue life by comparing LPBF and electron beam melting (EBM) technologies. Vrancken *et al.*[13] have found that the transformation of martensitic microstructure and variations of mechanical properties of Ti64 depend on post heat treatment. The fundamental findings from these studies still remain to be translated to optimization strategies for the LPBF processes to achieve better material properties.

It would be ideal to have the capability to control the LPBF-printed material properties by controlling machine process parameters in the printing process. However, with the diversity of LPBF machines and raw material powders available and the uncertainties related to the operating conditions (e.g., laser stability and powder contamination), it is virtually impossible to establish a universal correlation between process parameters and material properties. In addition, it is difficult to benchmark a process with respect to certain properties, such as fatigue tests, since the relevant tests require a large number of samples and are time-consuming. On the other hand, non-intrusive characterization of porous structures can be done in a much faster manner and consumes much less resources. If one can establish a robust correlation between the porosity of the LPBF-printed materials and their mechanical properties, it is then possible to use non-intrusive characterizations, such as a computer tomography (CT) scan, to evaluate if the part is acceptable. Such correlations have been studied in the literature using physics-based modeling techniques[9,14,15]. However, such models also demand high computational resources and their accuracy depends on input parameters (e.g., melt pool size, crystalline orientation), which usually require extensive experiments to determine. Moreover, the integration of pores into such simulations is not trivial as this is limited by the finite domain size the model can handle. It is thus desirable to have simple surrogate models to avoid the above-mentioned challenges, and data-driven models can be a potential solution.

To address some of the above-mentioned challenges, in this study, we investigated the effects of surface roughness and pore characteristics of LPBF-printed Ti64 parts on their fatigue lives and established data-driven surrogate models for their relationships. 197 fatigue bars were printed using a metal AM system (3D Systems ProX 320™) with varying the laser scanning speed, which altered the local melting and fusion process of the powder and in turn led to variable porous structures. The pores were then characterized using micro-CT. To investigate the surface roughness effect, half of the samples were grit-blasted, and the other half were machine-finished, which are characterized using optical surface profilometry. The statistics of the micro-pore density, location, size and shape, as well as surface roughness were systematically analyzed. Selected samples were then mechanically tested for their fatigue properties. The correlations between surface and pore features and fatigue properties were analyzed. Finally, a machine learning model using a drop-out neural network (DONN) was trained to link the porosity and surface roughness to the fatigue data. Besides predicting fatigue life, the DONN also has the unique capability of estimating the prediction uncertainty. The evaluation of fatigue life given pore and surface input data only take a few seconds using this DONN model. Practically, such efficient surrogate models may serve to reduce the amount of physical testing needed for LPBF-generated components by informing the user if the property of a printed component is within acceptance limits.

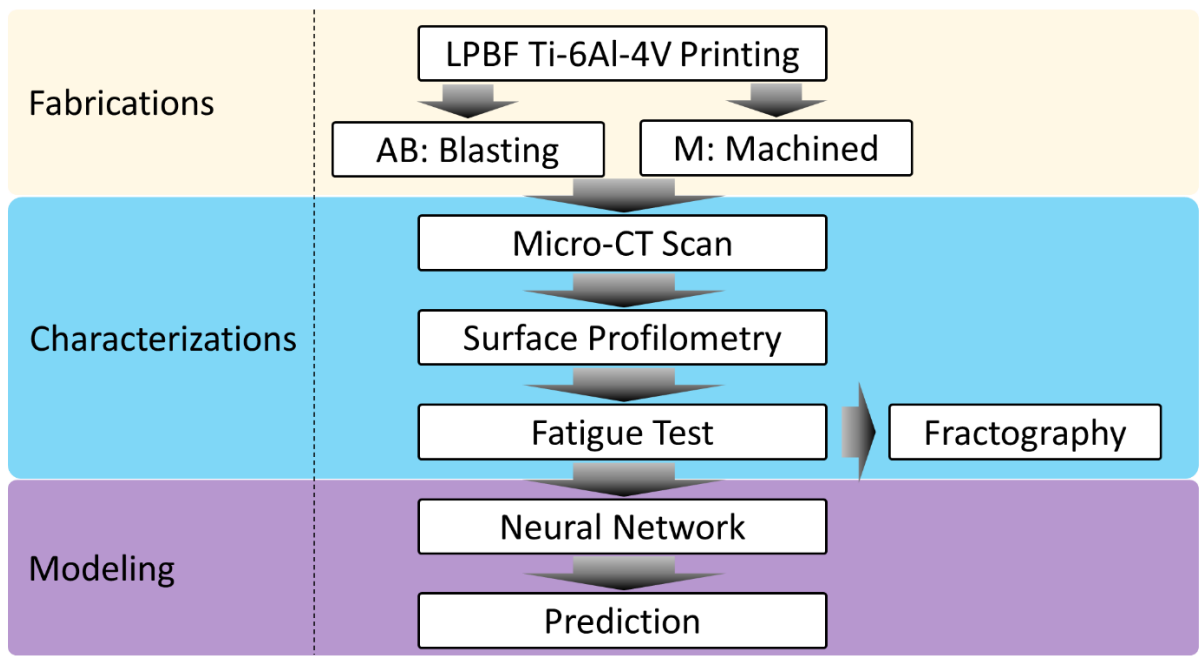

**Figure 1.** Scheme of this study.

## 2. Methods

### 2.1. Sample Printing

As described above 197 fatigue bars (21.08 mm × 2.54 mm × 84.58 mm) compliant with ASTM test methods were printed using a ProX DMP 320 AM system. 72 samples were used for fatigue life tests, 12 samples were utilized for CT scans, and 113 samples were scarified for preliminary tests (e.g., repeatability, laser power, speed and hatch distance tests). Ti64 metal powder (3D Systems LaserForm® Ti Gr23 (A) powder), which is of critical importance to a wide range of applications, like many aerospace and orthopedic components, was the material used. Different processing parameters including laser power, scanning speed, hatch distance and surface finish were varied so that the porous structures could be tuned in the preliminary porosity investigation (see supplementary material, **Fig. S1**). All parameters were varied by up to +/-20% of the machine recommended value. Optical micrographs of polished sample cross-sections were taken to examine the change in the pore statistics. It was found that among all of the processing parameters that were varied, scanning speed led had the greatest impact on pore density variation (**Fig. 2**)[16,17]. This preliminary study helps us identify the most effective process parameter to tune the internal structures of the LPBF parts. Since the purpose of this study was to link porosity, instead of processing conditions, to fatigue life, scanning speed was selected as the independent variable to be systematically varied in the printing process since it offers a wide range of micro-pore variations. To this end, the laser scanning speed was varied from 750 to 2000 (mm/s) with a 250 increment, where the vendor-recommended speed was 1250 mm/s.

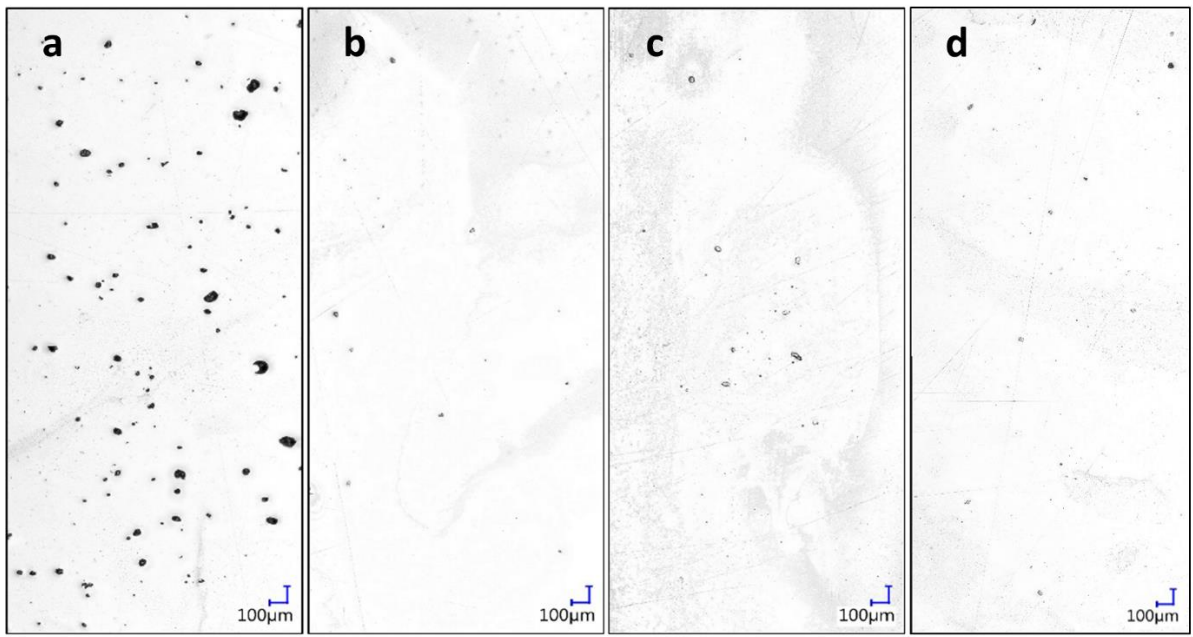

**Figure 2.** Micrographs of selected cross-sections showing pore density resulting from different printing

speeds. Laser scanning speed of **(a)** 750 mm/s (porosity area ratio (PAR): 1.29%), **(b)** 1250 mm/s (PAR: 0.05%), **(c)** 1750 mm/s (PAR: 0.14%) and **(d)** 2000 mm/s (PAR: 0.37%).

The printed samples were then heat-treated to release residual stress[10,13]. The samples were enclosed in the vacuum chamber and the heat treatments were executed at 650 °C for 2 hours. Then, the samples were divided into two groups, with one group machined (M) and the other as-built (AB) but grit-blasted. The abrasive grit blasting is a surface treatment process to remove the loose adhering powder. 120 grit aluminum oxide grains are accelerated through a blasting nozzle by means of compressed air. This yields different surface finishes and thus different surface roughness, another parameter that can potentially impact fatigue life besides internal micro-pore structures. We note that the machined samples were printed with a slightly larger thickness (0.5 mm) so that after machining, the dimension is the same as that of the grit-blasted sample. The printed samples were then machined into dog-bone geometry for fatigue failure testing.

The scan strategy in this experiment first used two contour scans offset from each other by 70 μm followed by the interior hatching scans[18]. The contour parameters were fixed for all samples produced. The rotation angle between layers was 245°. The surface roughness of the printed samples can potentially change due to the following three reasons. First, the melt pool changes depending on the laser speed, causing morphology changes[19]. The surface roughness reflects the rugged solidification of the melt pool as shown in **Fig. 3**. Secondly, the change in pore density and geometry caused by the laser speed as illustrated in **Fig. 2**. The change in porosity area ratio (PAR) by different laser speeds varies from 0.05 % to 1.29 %, and the relatively higher pore density can affect the surface roughness by showing open pores on the top of the printed surface[16,20]. Finally, the powder layer thickness can strongly impact the surface roughness that leads to unstable melt flow due to increased misalignment of the laser scanned tracks[16]. To avoid such variation, the thickness of the powder layer was fixed at 60 μm. These factors may impact the M or AB samples differently. Many researchers reported that surface roughness has a significant effect on fatigue crack initiation[7,8,12,21]. However, in terms of porosity, Tammas-Williams et al. insisted that the internal pores appear to have less of an effect on the mechanical property of printed samples[9].

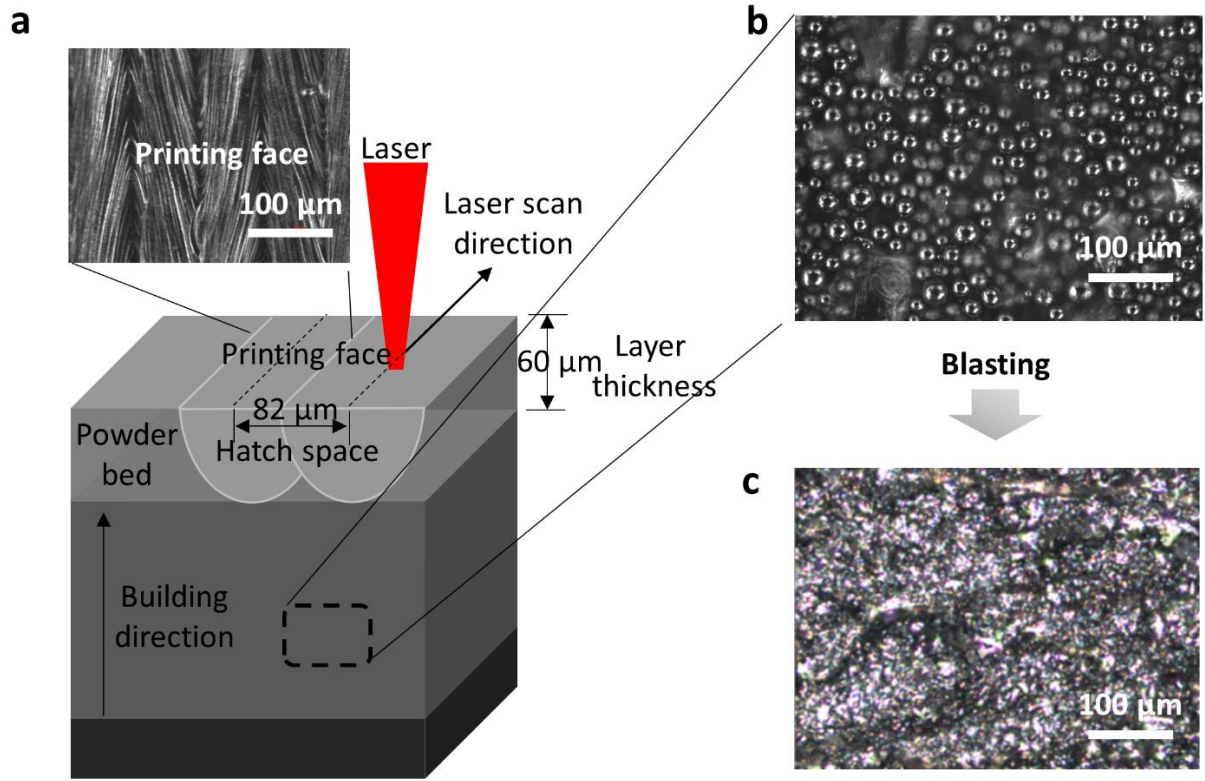

**Figure 3.** **(a)** Schematic LPBF process using Ti64 powder (inset: picture of printing face). **(b)** The surface of as-printed sample, where un-melted metal powder particles are loosely attached on the surface. **(c)** After blasting the surface, the loosely attached microparticles are removed. The surface roughness measurements are conducted on the blasted surfaces for the as-built samples.

Technically, the raster hatch scanning method in LPBF will not make the inner and surface pore features different. In this study, we used two contour scans prior to the hatch scan to impose different heating histories of materials close to the border of the printed sample from that of the internal materials. In this way, the surface and internal pore features become different, so that we have a way to study their impacts on fatigue life. **Fig. 4** shows the vertically built samples and the specific scanning path used. The two contour scans as shown in **Fig. 4b** were able to produce a depletion zone of pores so as to isolate the internal and surface pores (See **Fig. 9**).

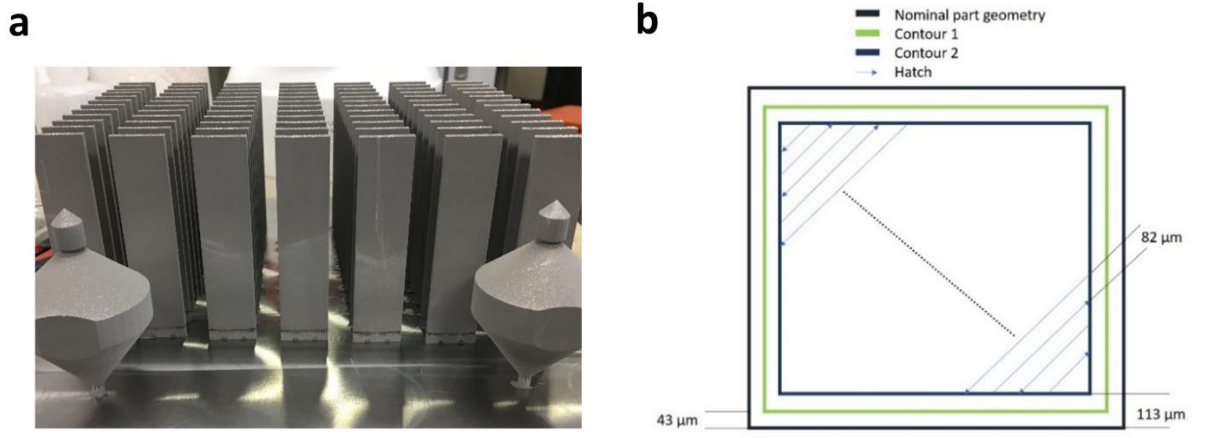

**Figure 4. (a)** Vertically built samples, which have an as-built dimension of 21.08 mm × 2.54 mm × 84.58 mm. **(b)** Printing strategy: the figure shows the cross-section for the vertically built sample. The hatch orientation rotates 245 degrees between layers. The black line is the as-designed part geometry. This line does not get scanned. Instead, the machine offsets the contours so that the edge of the contour 1 (green line) melt pool sits on the nominal part geometry. The order of scanning is contour 1, contour 2 and hatching. Here, the scan speed for the contour 1 and contour 2 were constant at 3000 mm/s. But hatching speed varied from 750 mm/s to 2000 mm/s.

## 2.2. Sample Characterization

The printed samples were then shaped into fatigue bars (**Fig. 5a** and **5b**) and subjected to various characterizations. Micro-CT (North Star Imaging X7000 system) was used to scan the internal pore features, and optical profilometry (Olympus LEXT OLS4100 confocal microscope) was used to characterize the surface roughness. Fractography using an optical microscope followed by scanning electron microscopy (SEM) was used to further understand the crack initiation of selected fatigue-tested samples.

**Micro-CT:** A micro-CT machine (North Star Imaging X7000 system) was used to characterize the pore features non-destructively. The equipment is capable of detecting pores with voxel size above 14 µm. More accurate measurements are possible, but the higher nominal resolution is coupled with a longer scanning time and yields a larger amount of data. The pore size detection based on the current resolution would be 28 ~ 42 μm. The whole gauge region was scanned, and pore features were collected. The VGSTUDIO MAX 3.3 Cast & Mold Extended software recorded the total number of pores, and for each detected pore, the coordinate, diameter, compactness and sphericity were calculated. Selected reconstructed micro-CT scans are shown in **Fig. 5**. The statistics of these pore features were then analyzed and quantified, which were later used to analyze fatigue failure and as inputs for the DONN.

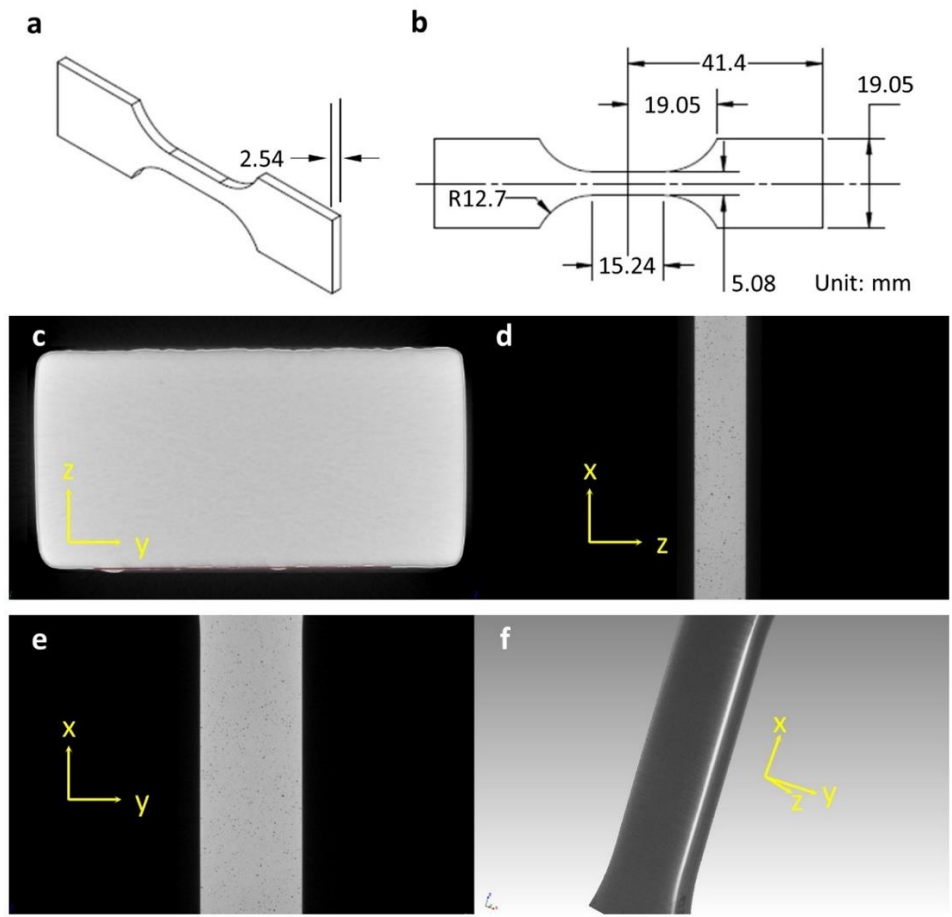

**Figure 5. (a)** The geometry of the fatigue bar. The unit is mm. The specimen profile was cut by wire electrical discharge machining (EDM) followed by longitudinal polishing of the edges to a surface finish of $R_a$ = 0.4 μm. Machined (M) specimens had faces machined by low-stress grinding followed by longitudinal polishing to a surface finish of $R_a$ = 0.4 μm. **(b)** Dimensioned drawing of the fatigue bar. **(c)**

Representative reconstruction of a micro-CT scanned sample. Images of laser scanning speed 750 mm/s sample for cross-section. **(d)** Side view. **(e)** Top view. **(f)** Three-dimensional view. Note the build direction is along the x-direction.

**Surface Profilometry:** For each sample, an optical profilometer (Olympus LEXT OLS4100 confocal microscope) was used to measure the surface roughness. Samples with two different surface finish methods, AB and M samples, were characterized. For each roughness data point reported, it is calculated from 20 different line profiles with the error bar representing the standard deviation. **Fig. 6** shows the representative surface profiles. Surface roughness parameters including mean roughness ($R_a$), maximum peak-to-valley roughness ($R_t$), 10-point height roughness ($R_{iso}$) and average radius of curvature of the deepest valleys ($\bar{r}$) were characterized from line scans (**Fig. 6b** and **6c**)[22]:

$$R_a = \frac{1}{l}\int_0^l |y| \tag{1}$$

$$R_t = |y_{max} - y_{min}| \tag{2}$$

$$R_{iso} = \frac{1}{5}\left[\sum_{i=1}^{5}(y_i)_{max} + \sum_{i=1}^{5}|(y_i)_{min}|\right] \tag{3}$$

$$\bar{r} = \frac{1}{5}\left[\sum_{i=1}^{5}(r_i)_{min}\right] \tag{4}$$

where $y$ is the height of line profile, $y_{max}$ is the maximum peak, $y_{min}$ is the minimum valley and $r_i$ is the radius of the deepest valley.

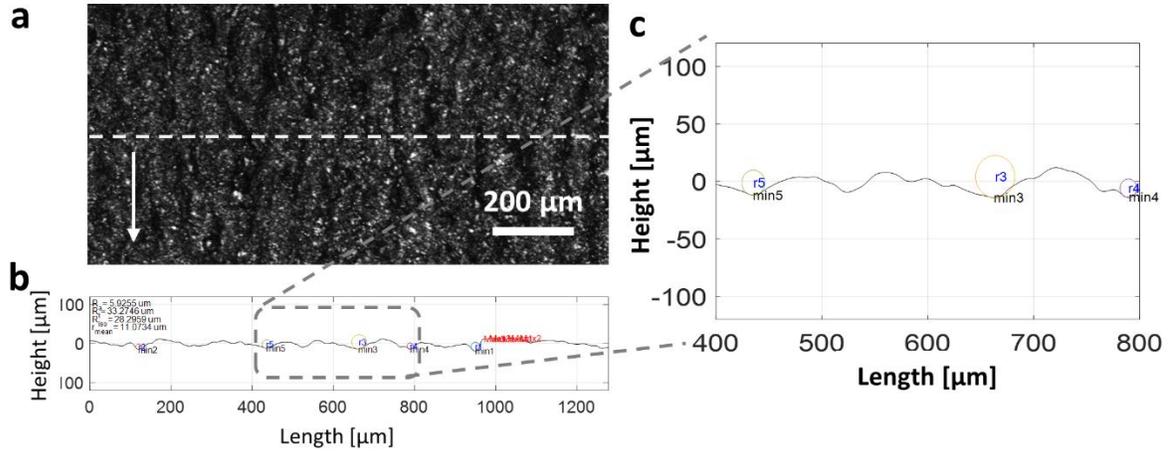

**Figure 6. (a)** The surface of a typical sample built at laser speed 1250 mm/s. The surface image was taken after blasting. **(b)** The line profile of the dashed line in (a). **(c)** Enlarged line profile of the selected area in (b). The circles represent the radius of the curvature in the deepest valley.

### 2.3. Stress-controlled Fatigue Testing

The stress-controlled fatigue test per ASTM E466 was performed with an extensometer[23]. Stress-controlled fatigue is considered to be applicable in cases where the strains are predominately elastic. We monitored strain using an extensometer and observed very limited plasticity, even in the highest stress level tests. Thus, stress-controlled fatigue tests per ASTM E466-15 were conducted. Fatigue behaviors of the samples were measured using load-controlled axial fatigue testing at room temperature. Unidirectional stress (stress ratio = 0) tests were performed with the range from 414 MPa to 1034 MPa. Trapezoidal loading waveform with a frequency of 15 cycles per minute (CPM) was used for the fatigue tests. The fatigue test at about maximum stress of 552 MPa that reached $10^6$ cycles without failure was treated as runout. A complete fracture within the gauge section of the test sample was considered as a failure.

### 2.4. Fractography

A fractography analysis was performed to characterize the fatigue failure. The fracture origins were visually examined by low magnification of a stereo microscope (Meiji Techno) under white light illumination. The detailed evaluation was performed by a field emission SEM (Magellan 400, FEI). The entire fracture surfaces were examined in this evaluation and if the fracture origins were identified, the information of the origins such as pore locations and size were documented.

## 2.5. Dropout Neural Network (DONN)

DONN [24] is a machine learning model that can be used as a surrogate model in regression tasks and at the same time capture model uncertainty. It has been proven to be equivalent to Bayesian neural network (BNN), which also produces model uncertainty besides predicting results, but DONN is much easier for implementation[24]. In addition, the main reason for choosing DONN over BNN is that the former is much less computationally expensive, especially as the data size scales up. Thus, the advantage of DONN will stand out more obviously when dealing with large amounts of information, which is expected to be the case as more data become available in the future. After training, evaluation using DONN only takes a few seconds. **Fig. 7** below shows both a standard neural network and a DONN. With dropout, binary variables for every input point and for every network unit in each layer (except the last one) are sampled, and each binary variable takes value 1 or 0 with a predefined probability for each layer. A unit will be dropped (i.e., its value is set to zero) for a given input if its corresponding binary variable takes the value 0. We use the same values in the backward pass propagating the derivatives to the parameters. For example, if 40% binary variables take values 0 in the forward process, then 40% binary variables will take values 0 in the backward process, so that only part of the parameters will be updated in the backward process. When training a standard neural network with dropout techniques, it can be regarded as training an ensemble of neural networks at the same time. When the training is finished, we can perform stochastic forward passes through the network with dropout applied to obtain the prediction distribution, where the average prediction and standard deviation (uncertainty) can be calculated.

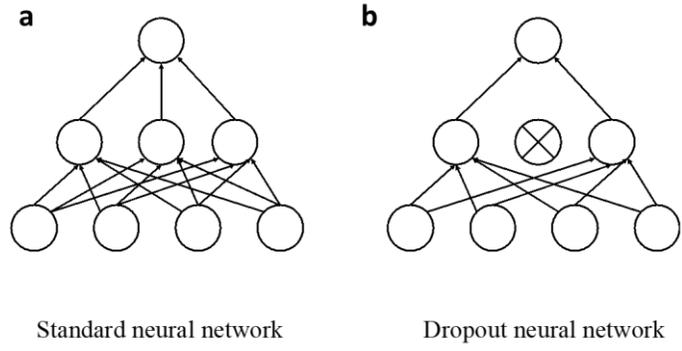

**Figure 7.** Schematics of a standard neural network **(a)** and the DONN **(b)**.

## 3. Results and Discussion

### 3.1. Surface Roughness

**Fig. 8** shows the surface roughness parameters as the laser speed increases. In the cases of AB samples, un-melted metal powders are attached to the surface parallel to the laser beam (see **Fig. 3**)[22]. These micro-sized powders tend to detach easily during the measurement of surface roughness. These features interfere with obtaining reliable measurement values. Thus, we examine the surface of the samples after the blasting process for the AB samples.

As shown in **Fig. 4b**, the surface regions of all AB samples were built by the double contour scans with constant speed (3000 mm/s). Thus, the variations of the hatching speed do not affect the surface roughness. However, the internal porosity changes when the hatching speed changes. Thus, as shown in **Fig. 8**, there is no clear relationship between laser speed and roughness in the cases of AB samples. This is because the surface pores were controlled through the dual contour scans, and high laser speed of 3000 mm/s, which dramatically increases the porosity level (**Fig. 2d**), was excluded from our experiments. Qiu et al. also reported the uniform roughness of printed surfaces when the porosity level is relatively low[16]. Our surface inspection results of the blasted surfaces were consistent with their results at low porosity level.

The surface of each M sample was polished along the longitudinal axis (x-direction in **Fig. 5**) to have almost constant average surface roughness ($R_a = 0.4 \pm 0.1$ μm) so that the roughness effect on fatigue behavior can be restricted. However, the two groups, AB and M samples, have noticeably different surface conditions regardless of laser speed. We should note here that both groups have almost constant roughness parameter values, except one outlier at 1000 mm/s laser speed. The reason is as follows. **Fig. 8d** represents the average radius of curvature at the deepest valleys. Thus, in the case of the AB samples which have relatively large $R_a$ values, the average radii of curvature are almost constant because the surfaces of the AB samples are uneven (i.e., deep valleys). Conversely, if the surface is flat, the radius of curvature at the deepest valleys we have designated will be very large and random. As a result, in the case of flat surfaces, it does not necessarily guarantee constant values of the radius of the curvature. Therefore, the values of M samples in **Fig. 8d** are relatively high and random, implying the feature of flat surfaces.

In addition, in the case of $R_a$, $R_t$ and $R_{iso}$, the AB samples had much higher values than M samples, but the average radius of the curvatures at the lowest valleys showed the opposite trend. The reason is that a larger radius of curvature is calculated on a relatively slowly varying surface, where shallow micro-notches have larger radii of curvature. Consequently, when surface roughness indeed affects the initiation of the cracks during the fatigue tests, we should be able to see the distinguishable characteristics of both groups: having the same pore features but different surface conditions.

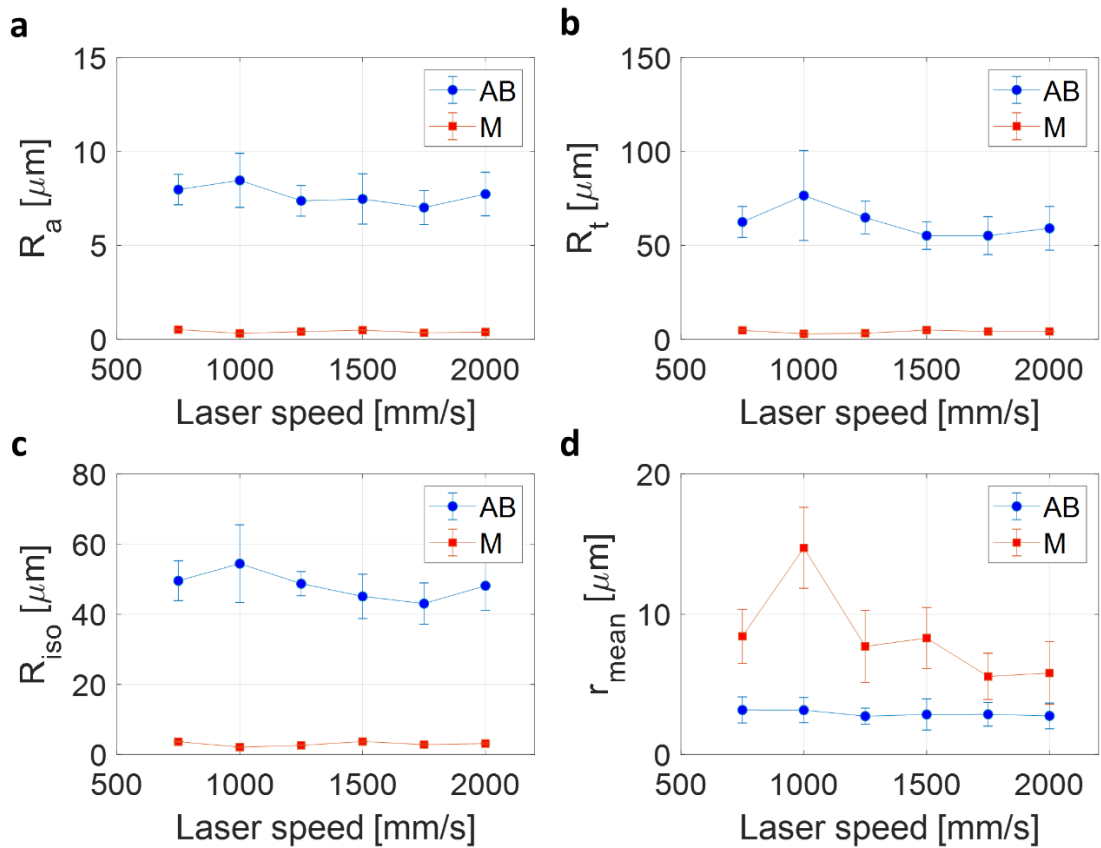

**Figure 8.** Surface roughness of blasted (AB) and machined (M) samples. Each data point is calculated from 20 different line profiles, and each error bar indicates the standard deviation for the 20 measurements. **(a)** Average roughness as a function of the laser speed. **(b)** The maximum peak-to-valley roughness. **(c)** 10-point height roughness. **(d)** Estimated average radius of the curvatures at the lowest valleys.

### 3.2. Pore Characteristics

In the case of the AB samples, the pores can be readily divided into two groups: internal and surface pores. The two groups are isolated by a depletion zone created by the contour scans (**Fig. 4b**), which re-melt the location to minimize pore formation. The destined hatch lines (**Fig. 4b**) are scanned by actually extending the hatch lines past where they are supposed to end, but turn off laser at the end of each hatch line. In the same vein, when starting a new line scan, the actual starting point of the hatch scan is the outside of the part exposed with no power, but the laser turns on at the starting point of the destined hatch line. Thus, the contour 2 line gets melted twice; once by contour 2 and the other by hatch passes. Such re-melting should be the cause of the depletion zone.

**Fig. 9** shows that the features of the surface and internal pores are distinguishable in terms of locations, shape and dimension. Usually, internal pores are formed due to the insoluble gas bubbles trapped during solidification, keyhole induced porosity and lack of fusion voids[15,25-27]. The relation between the internal pore volume (measured in *voxel*) and diameter follows a power law of 2.1, while that of the surface pores is much smaller at 1.5 (**Fig. 9b**). This finding suggests that the surface pores are farther away from a spherical shape (i.e., more irregular) than the internal pores. This is also supported by **Fig. 9c**, which shows that the surface pores exhibit a different sphericity-compactness relation than internal pores. This is further supported by the much larger disparity in the projected areas on the XY- and YZ-planes of the surface pores than the internal pores. By carefully examining the CT scan, it is evident that the irregular shape is caused by the open pore structures exposed to the surface (**Fig. 9e**). In particular, among the many pore features, the projected pore area normal to the applied stress direction during fatigue test is considered to be a key factor of crack initiation[9,28]. From that perspective, it is interesting the projected area of the surface pore on the XY-plane (parallel to the sample surface) is larger than the projected area on the YZ-plane (normal to applied stress direction) due to widely opened structures (**Fig. 9d**).

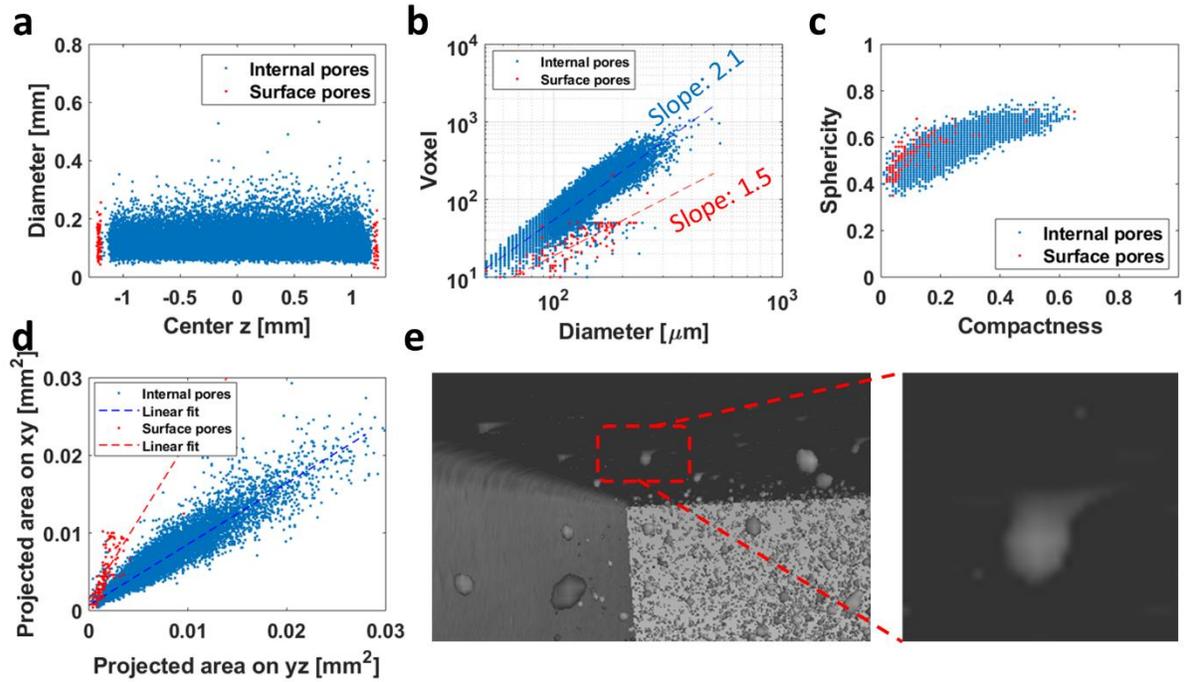

**Figure 9.** Pore features for AB samples with laser speed 750 mm/s. **(a)** A typical example of depletion zone created by the contour scans. The data show the location (z-coordinate, see **Fig. 5**) and the diameter of the pores extracted from the CT scan of a 750 mm/s speed sample. Surface pores are defined as those within ~80 μm from the sample boundary, where the diameter is the circumscribed sphere of the pore. More examples for the depletion zone are presented in **Fig. S2**. **(b)** A plot of the voxel of pores as a function of their diameter, where the voxel represents the number of voxels a pore has from the CT data. **(c)** A scatter plot of compactness vs. sphericity of the pores. **(d)** Projected areas of the pores. **(e)** Reconstructed CT images of pores. The inset reveals the open feature of a typical surface pore.

As mentioned previously, the M samples were printed with a larger thickness in the z-direction (see **Fig. 5**) than AB samples, since the M samples are to be polished to the same dimension as the AB samples. About 200 μm in thickness was removed for planarization for these samples, meaning that the depletion zone was removed and internal pores exposed. The exposed surface pores by the polishing process were different from the surface pores of the AB samples. The features for the M samples are displayed in **Fig.**

**10** and the surface pores for these samples are defined as those within ~80 μm from the sample's polished surface. Since the surface pores of the M samples can also be cut-off by the polishing process, they can show different features compared to the internal pores as shown in **Fig. 10b-d**, but the differences are much smaller than those in the AB samples (**Fig. 9**). In particular, **Fig. 10b** and **10d** show that the voxel and projected area are reduced by the cut-off effect, but **Fig. 10c**, which shows the same sphericity-compactness for the internal and surface pores, indicates they are of the same origin. Here, we note that the distribution of "cut-off surface pores" by polishing process (**Fig. 13a**) is different from "opened surface pores" as shown in **Fig. 9d**. We also note that the densities of the exposed pores are too small to influence surface roughness of the M samples, which is evident in **Fig. 8**. From **Fig. 14** and **Fig. 15** in sections 3.5 and 3.6 of this manuscript, it can be seen that the average pore features are highly correlated for M samples while decoupled for AB samples. The detailed correlations will be discussed with fatigue behaviors in Section 3.5 and 3.6.

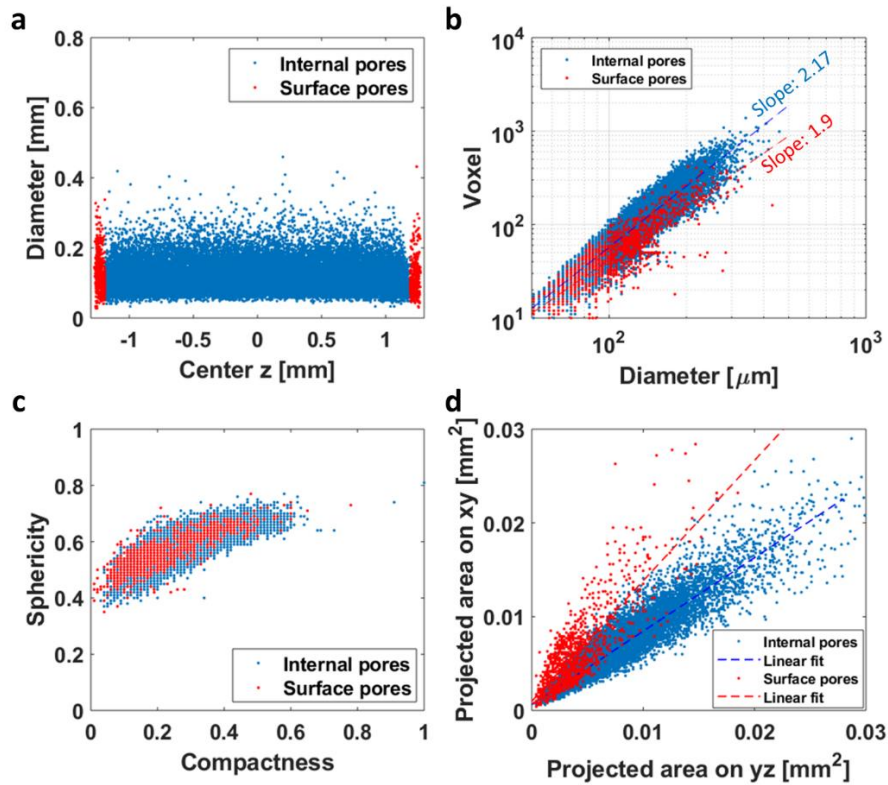

**Figure 10.** Pore features for M samples with laser speed 750 mm/s. **(a)** Scattered plot of diameter as a function of z-coordinate for M750. Surface pores are defined as those within ~80 μm from the physical boundary. **(b)** A plot of the voxel of pores as a function of their diameter for M750. **(c)** A scatter plot of compactness vs. sphericity of the pores for M750. **(d)** Projected areas of the pores for M750.

### 3.3. Fatigue Test Results

**Figure 11** shows the difference in fatigue life for the two groups of samples, AB and M samples, with varying printing speeds. Regardless of the laser speed, AB samples exhibit a relatively narrow distribution in the S-N (Wöhler) diagram (**Fig. 11a**), which is likely due to the large surface roughness (**Fig. 8**) dominating the low-cycle fatigue (LCF)[29]. The effect of the inner pore is comparatively small when the effect of surface roughness is dominant. On the other hand, the M samples, of which the surface roughness effect is expected to be small, show relatively wider distributions in the LCF data (**Fig. 11b**) compared to the AB samples. This is because the internal pores are exposed to the surface during the polishing process, so the influence of the porosity effect or other parameters, especially that can be varied by the laser speed, on the fatigue life is more obvious than the AB samples. It is worth noting that the data of the M2000 sample (i.e., machined sample printed with a laser speed of 2000 mm/s) records the lowest fatigue life although these samples have significantly lower internal pore density than M750 or M1000 samples as shown in **Table 1**. Therefore, the most detrimental influence on the M samples is not the inner pore density, but according to **Table 1**, the pore size and the projected area of pores normal to the applied stress, which is consistent with findings from Ref.[9,28]. The detailed correlations will be discussed in Section 3.5.

Classically, the fatigue life prediction is based on the Basquin power law which is represented by the following equation[30,31]:

$$\sigma = cN^m \quad (5)$$

where $\sigma$ is stress amplitude, $N$ is number of cycles to failure and $c$ and $m$ are the fitting parameters in the Basquin's model. However, often linear-logarithmic coordinate was adopted to describe the experimental S-N data [7,32-34]. The linear-log form in the finite life region is given by:

$$\sigma = a(logN) + b \tag{6}$$

where *a* and *b* are the fitting parameters of the linear-log form. The validity of this fatigue model was tested by taking into account the determination coefficient ($R^2$) of each fitting function. This validation is critical because proper data should be fed to the later machine learning for training. As shown in **Fig. 11c**, the linear-log model shows a high level of agreement ($R^2$ above 0.91) in the low-cycle fatigue region selected at $N < 10^5$. Therefore, the linear-log data was adopted in the preset work for fatigue analyses and the DONN model construction.

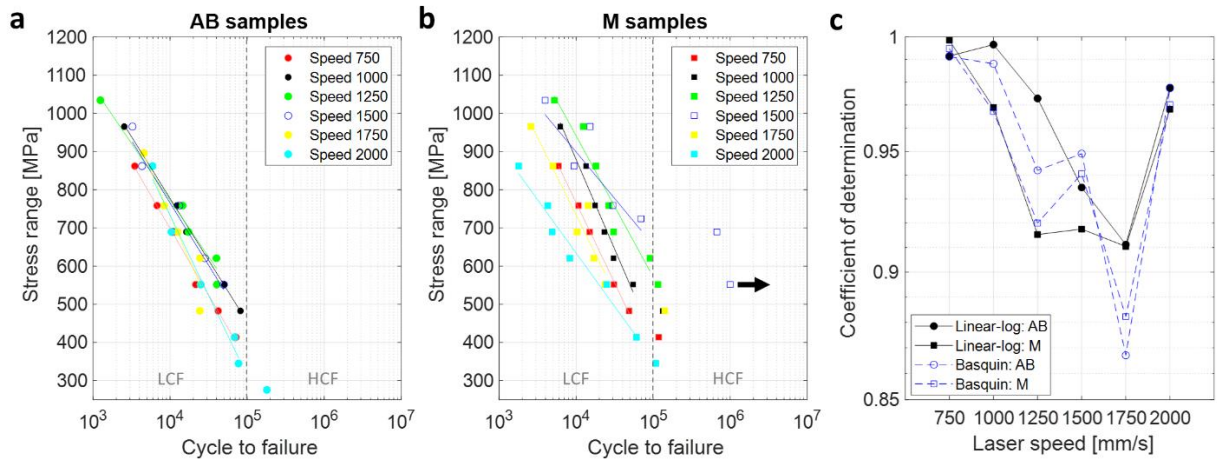

**Figure 11.** **(a)** The plot of stress-fatigue life (S-N) for AB samples. Conventionally, the S-N plot is partitioned into low-cycle fatigue (LCF) and high-cycle fatigue (HCF) regions at $10^5$ cycles [33]. The fitting results from the linear-log model included the LCF region of data. **(b)** The S-N plot for the machined (M) samples. The fitting results from the linear-log model also included the LCF region of data. The arrow marker represents the runout. **(c)** Comparison of the linear-log and Basquin models based on the $R^2$ of each fitting function. The fitting lines in the log-log coordination were also presented in the supplementary information (**Fig. S3** and **Table S1**).

**Table 1.** Data comparison using the M samples to find key factor for the fatigue life. IP and SP refer to internal pores and surface pores. The projected area of the pore on YZ-plane is denoted as $\sqrt{Area}$ in μm unit.

| Laser Speed | Surface Condition | IP # density (mm$^{-3}$) | SP # density (mm$^{-3}$) | IP sum voxel | SP sum voxel | IP mean voxel | SP mean voxel | IP $\sqrt{Area}$ | SP $\sqrt{Area}$ | LCF cycle @ 758 MPa |
|---|---|---|---|---|---|---|---|---|---|---|
| 750 | Machined /Polished | 195.6 | 89.6 | 3469713 | 80167 | 95.9 | 69.6 | 76.2 | 63 | 10729 |
| 1000 | Machined /Polished | 39.9 | 8.9 | 678475 | 5070 | 91.8 | 44.5 | 74.1 | 50.3 | 17656 |
| 2000 | Machined /Polished | 5.1 | 3.3 | 258997 | 3719 | 236.2 | 63.0 | 114.8 | 60.1 | 4283 |

### 3.4. Crack Initiation Analysis

Fracture surfaces are comprised of three different zones: origin of crack initiation, crack propagation and finial overload zone [35]. The critical factors for the fracture phenomenon are associated with surface roughness [7,8] and porosity [9,35]. However, fracture is often a cross-correlated process, making it usually difficult to design a model that can draw simple and clear conclusions. Thus, we use SEM to identify some common features of the fractured surfaces. Since the AB samples, the surface roughness-dominant group, display a relatively narrow distribution in the S-N curve regardless of laser speed changes, we can speculate that cracks initiate from unfilled surface cavities [12]. Except for the AB0750 sample, the cracks of the other AB samples (**Fig. 12b, 12c** and **12d**) all initiate from the surface, which is expected. One exception is the AB0750 sample (**Fig. 12a**), which is pore-rich, has the crack originated from a pore located just beneath the surface.

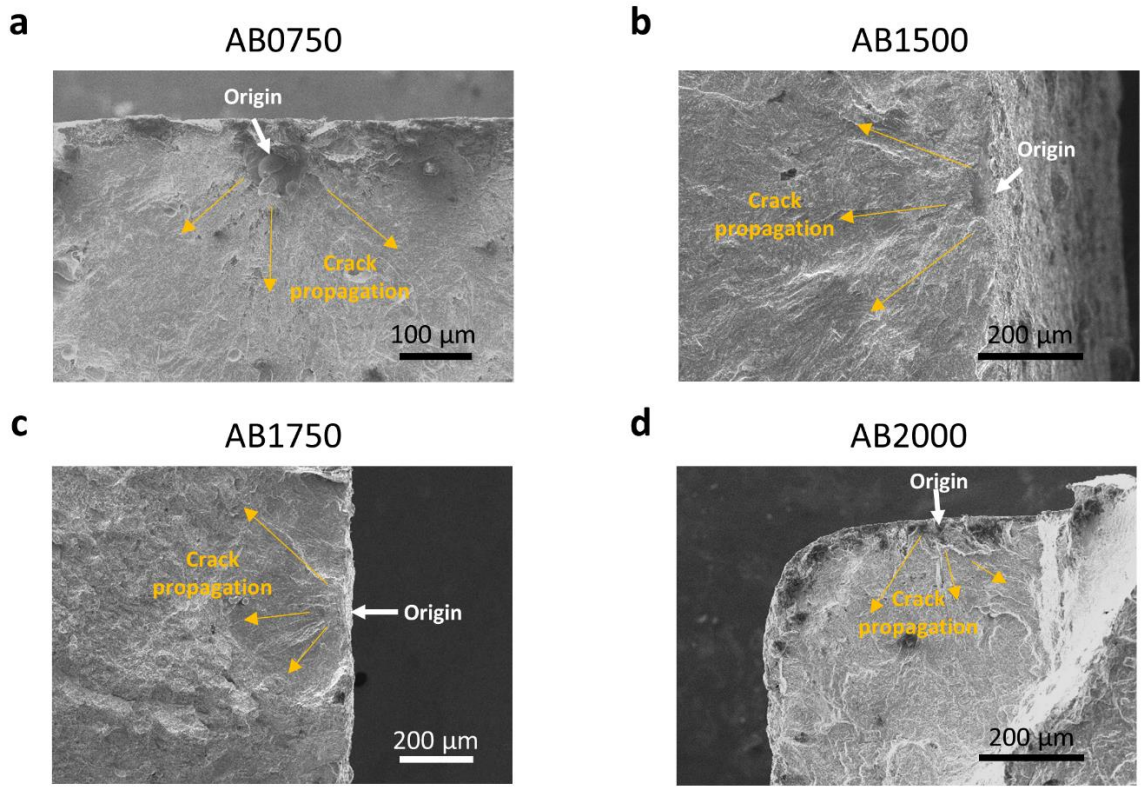

**Figure 12.** Fractography SEM images of AB samples. (a) For an AB750 sample (as-built sample printed at a speed of 750 mm/s), crack initiates from a pore located on the sub-surface. (b) For an AB1500 sample, crack initiates from the surface. (c) For an AB1750 sample, crack initiates from the surface. (d) For an AB2000 sample, crack initiates from the surface.

It is necessary to focus on analyzing the crack initiation using the M samples, because the M samples are expected to be influenced by more convoluted pore features as the surface roughness is much smaller than the AB samples. For instance, crack initiation by LCF test can be related to the properties of micro-pores in the fatigue bars such as the size, location, and shape [28]. Especially, it is generally true that an excessively porous fatigue bar must have many pores on the surface concentrating stress around them. These surface pores are more likely to be crack initiators. Therefore, finding the crack initiations after the fatigue failure in our study is a way to account for how closely the estimation of statistical data correlates with the actual results.

In the case of the M0750 sample, five identifiable fracture origins were found, and all of them were from pores. Four of them were surface pores like **Fig. 13a** and one of the crack origins was located on the subsurface (within ~ 100 μm from the surface). In the case of samples with a low porosity, crack initiated from the surface or a fine defect at the corner (**Fig. 13b** and **13c**). For a M2000 sample, due to the fast scanning laser speed, relatively large pores were created and an irregular shape of the surface pore initiated a crack, which shows characteristics of lack-of-fusion (**Fig. 13d**) [20,36,37].

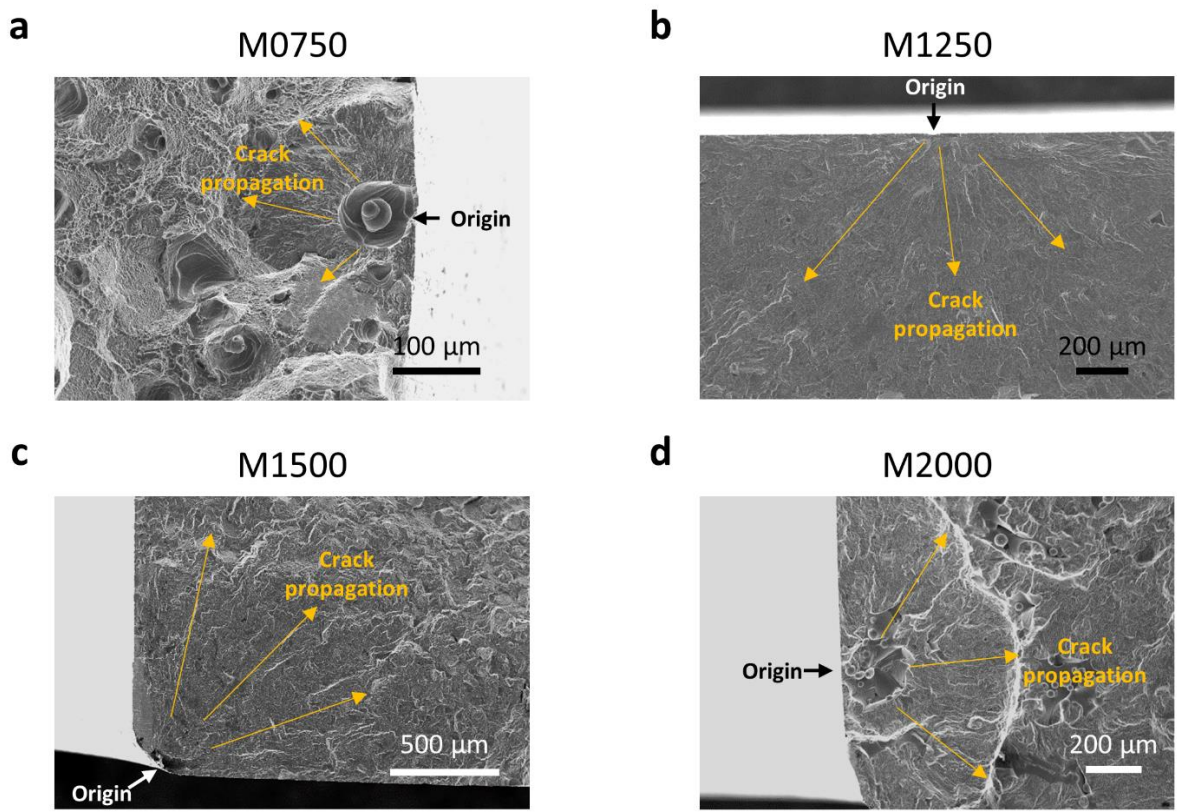

**Figure 13.** Fractography SEM images for various crack initiations. **(a)** For a M0750 sample (machined sample printed at a speed of 750 mm/s), crack initiates from an open pore located on the surface. **(b)** For a M1250 sample, crack initiates from the surface. **(c)** For a M1500 sample, crack initiates from the corner of the sample. **(d)** For a M2000 sample, crack initiates from a surface pore, which shows characteristics of lack-of-fusion.

### 3.5. Correlation between CT Data and Fatigue Life for M Samples

According to other studies [6-9], factors that can determine fatigue life include surface roughness and pore characteristics such as pore position, pore density and pore size. However, in our study, as mentioned in the Methods section, the M samples significantly reduced the surface roughness effect by the polishing process. Thus, it is necessary to take into account various parameters to examine fatigue life. For that reason, we conducted extensive statistical analysis of the CT data and LCF data.

**Figure 14** shows the relationship among fatigue life, various pore parameters and laser speed for the M samples. As the laser speed increases, pore number density tends to drop sharply at low speeds, but slowly increases when the speed is higher than 1500 mm/s (**Fig. 14b**). On the other hand, the volume size of the pore (mean voxel) also decreases first but increases rapidly after 1500 mm/s. The same trend can be seen in all other properties of the pores such as sum of voxel (i.e., total volume of pores, **Fig. 14c**) and projected area of pores (**Fig. 14d**). These suggest a transition from keyhole pores by locally excess power density due to long laser exposure time at low speed to lack-of-fusion pores due to insufficient heating/melting at high speed [37]. As can be seen from these analyses, it is important to note that for these pore features, the trends as a function of laser speed are the same for internal and surface pores, suggesting that they are of the same origin. In addition, what can be deduced from the results of the LCF and the printing speed is that the optimal condition for the printing speed is 1500 mm/s, different from the printer vendor recommended 1250 mm/s. However, even if the surface roughness effect is excluded for these M samples, it is still ambiguous as to what factors have the most significant effect on fatigue life because many pore characteristics parameters are correlated. We again emphasize that the M samples have highly correlated internal pore and surface pore features. In other words, even if we distinguish between the surface pores and the internal pores, the trend of the surface pores is dependent on the internal pores (as shown in **Fig. 14**), because the internal pores are exposed to the surface during polishing. The AB sample analysis where the surface and internal pores are decoupled from each other will be covered in the next section.

Pearson correlation coefficient (PCC) can present the quantified linear correlation between two variables. In **Fig. 14e**, the correlation coefficients for the M samples related to LCF cycles are displayed. For the log cycle to failure (*logN*) at the maximum stress 785 MPa, the average projected area of pores denoted as $\sqrt{Area}$ and the average size of pores, measured as mean voxel, show the strong negative correlations (-0.804 and -0.849 for internal pores, respectively) with LCF (**Fig. 14e**). The same observations are made for the other two analyzed stress levels. Although the pore number density and the sum of pore voxel inside the fatigue bar are related to mechanical strength (e.g., Young's modules and elongation)[38], the most critical parameter for fatigue life turns out to be the size of the pore normal to the applied stress. As expected, the PCCs for internal and surface pore features shows similar trends as seen in **Fig. 14e**.

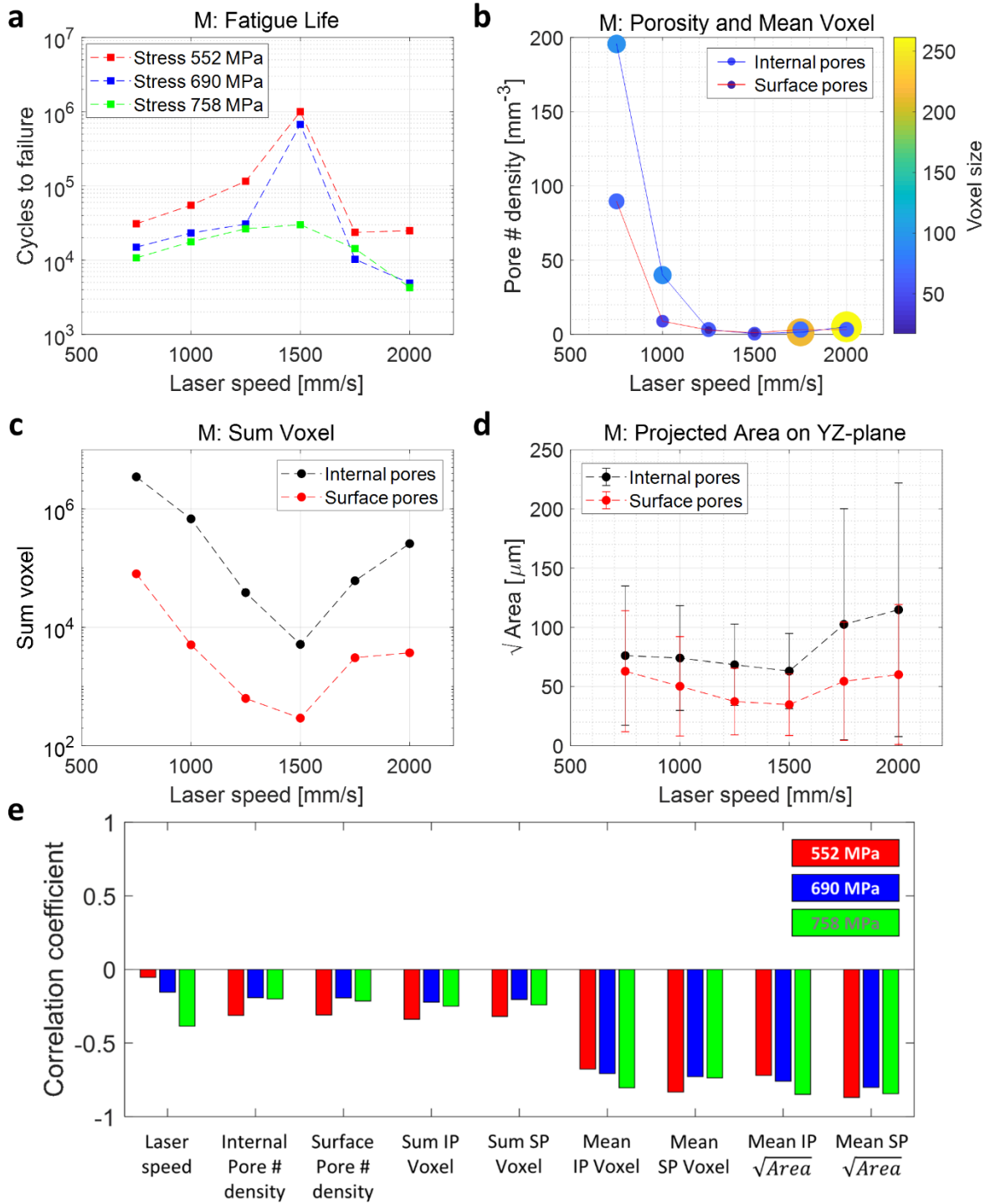

**Figure 14.** M samples data correlation between LCF results and pore characteristics as a function of laser speed. **(a)** Number of cycle to failure, **(b)** pore number density (size and color of the markers represent the mean voxel of the pore), **(c)** sum of pore voxel, and **(d)** projected area of pores on the YZ-plane (error bars

indicate the standard deviation), as a function of laser speed. **(e)** Correlation coefficients (M samples) related to LCF cycles (*logN*). The parameter for projected area on the YZ-plane is denoted as $\sqrt{Area}$.

### 3.6. Correlation between CT Data and Fatigue Life for AB Samples

The AB samples in general show comparatively shorter fatigue life than the M samples, which should be caused by different surface conditions between the two groups of samples (**Fig. 8**). Since the roughness of all AB samples is similar, the variation of the fatigue life is less affected by the laser speed compared to that observed in the M samples.

For the AB samples, the laser speeds of 1000 and 1250 mm/s generally lead to better fatigue lives regardless of the applied maximum stress (**Fig. 15a**) and the features of the internal pores as shown in **Fig. 15b**, **15c** and **15d** are almost identical to those of the M samples in **Fig. 14**. Unlike the surface pores, the features of the internal micro-pores are readily changed by the laser speed. At the relatively low laser speeds, a large number of pores are generated per unit volume, likely due to keyhole formation[19,37]. At relatively high laser speeds, the pore density is small, but the size of the micro-pores increases, likely due to lack of fusion. As shown in the case of M samples, we see that the optimized process condition to minimize internal pore density is established at 1500 mm/s from our observation, while the vender-recommended specification for the printing speed is 1250 mm/s.

However, AB samples have unique properties in terms of surface pores. For example, the surface pores, defined as those within ~80 μm from the sample's physical surface, have almost constant number of pore density and size regardless of the laser speed because the processing parameters of contour scans are fixed for all samples (**Fig. 15b**). For the average projected pore areas on the YZ-plane, the surface pores have smaller values than the internal ones (**Fig. 15d**). The correlation strengths between the pore features and LCF differ for internal and surface pores depending on the specific features we analyze. The LCF is more correlated with the density and total pore voxel of the surface pores than the internal pores as shown by the higher PCCs (**Fig. 15e**). For the mean voxel and projected area, surface and internal pores exhibit similar strength of correlation with LCF. The behavior can also be slightly different for different stress levels. For LCF at 758 MPa, the three largest coefficients are strongly related to the surface pore information (i.e., the sum of surface pore voxel: -0.993, surface pore number density: -0.987 and the mean surface pore voxel: -0.802). LCF at 552 MPa shows similar behavior, but behavior at 690 MPa is slightly different, likely due to the more irregular LCF data as shown in **Fig. 15a**. This observation implies that the management of the surface pores can have a larger impact on LCF than internal pores.

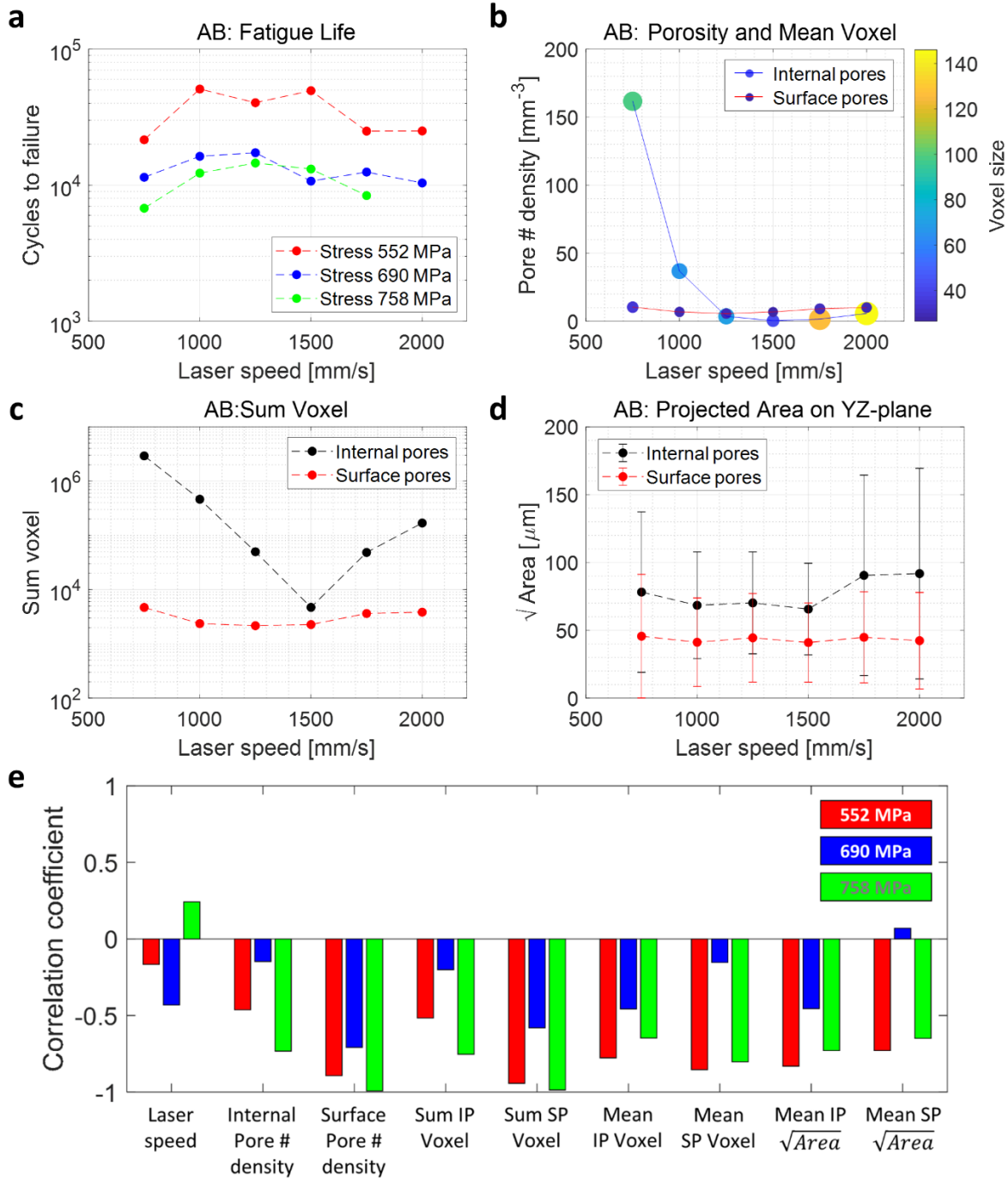

**Figure 15.** AB samples data correlation between LCF results and pore characteristics as a function of laser speed. (a) Number of cycle to failure, (b) pore number density (size and color of the markers represent the mean voxel of the pore), (c) sum of pore voxel, and (d) projected area of pores on the YZ-plane (error bars indicate the standard deviation), as a function of laser speed. (e) Correlation coefficients (AB samples)

related to LCF cycles (*logN*). The parameter for projected area on the YZ-plane is denoted as $\sqrt{\overline{Area}}$.

### 3.7. Drop-out Neural Network

We first quantify the relationships between the pore features and the log cycles to failure (*logN*) independently for the AB and M samples since they have very difference surface roughness. The descriptors used for M and AB samples are stress ($\sigma$), surface roughness (all four parameters: $R_a$, $R_t$, $R_{iso}$, $\bar{r}$), pore density ($\rho$), diameter ($\bar{d}$), compactness ($\eta$), sphericity ($\gamma$) and projected YZ area. Pore features for both internal and surface pores are included as independent descriptors. There are 41 LCF data points for the M samples and 35 for the AB samples, and we train the DONN using the leave-one-out cross-validation method (i.e., reserve one data for testing and use the rest data for training, which iterates through all the data) given the limited amount of data. The inputs and labels are all standardized before feeding into the DONN for training or validation. **Figure 16a** and **16b** respectively shows the pair plots between predictions from the trained DONN and the experimental values for M and AB samples. It can be seen that the models can predict the *logN* given a set of surface and pore descriptors with good accuracy. The DONN-predicted average *logN* agree well with the experimentally measured *logN*, with PCC of 0.935 and 0.944 respectively for the M and AB samples. It is noted that when PCC = 1, there is a perfect correlation between the prediction and ground truth. In addition, the prediction uncertainties are also shown, as color-coded in **Fig. 16**. It is seen that all of the prediction uncertainties are below 0.35 for the M samples (**Fig. 16a**) and below 0.13 for the AB samples (**Fig. 16b**).

Since the major difference between the AB and M samples are their surface features, we further trained a unified DONN using all data from both sets to predict *logN* of all samples. We then went through the same training process as the previous scenario and drew the pair plot between predictions and the experimental values in **Fig. 16c**. The high PCC value of 0.946 again indicates that the unified model still has good prediction capability, and the uncertainties are mostly below 0.2 with only one case of ~0.3. The reason of the high accuracy from DONN could be that the data collected from the experiment were of high quality, and the correlation between the pore features, surface roughness and fatigue life was well represented by the data collected, which was implied in **Fig. 11**. We have tested the DONN model with even less data points by randomly removing some from the database, but the DONN model still shows high predictive accuracy (see **Fig. S4**). It is possible that the model predictive capability may degrade if we are predicting pore features and surface roughness way out of the training range. However, the fact that DONN can be accurate and in the meantime estimate uncertainty suggests that such a model, with proper training against high-quality data, can be a useful tool for AM analyses.

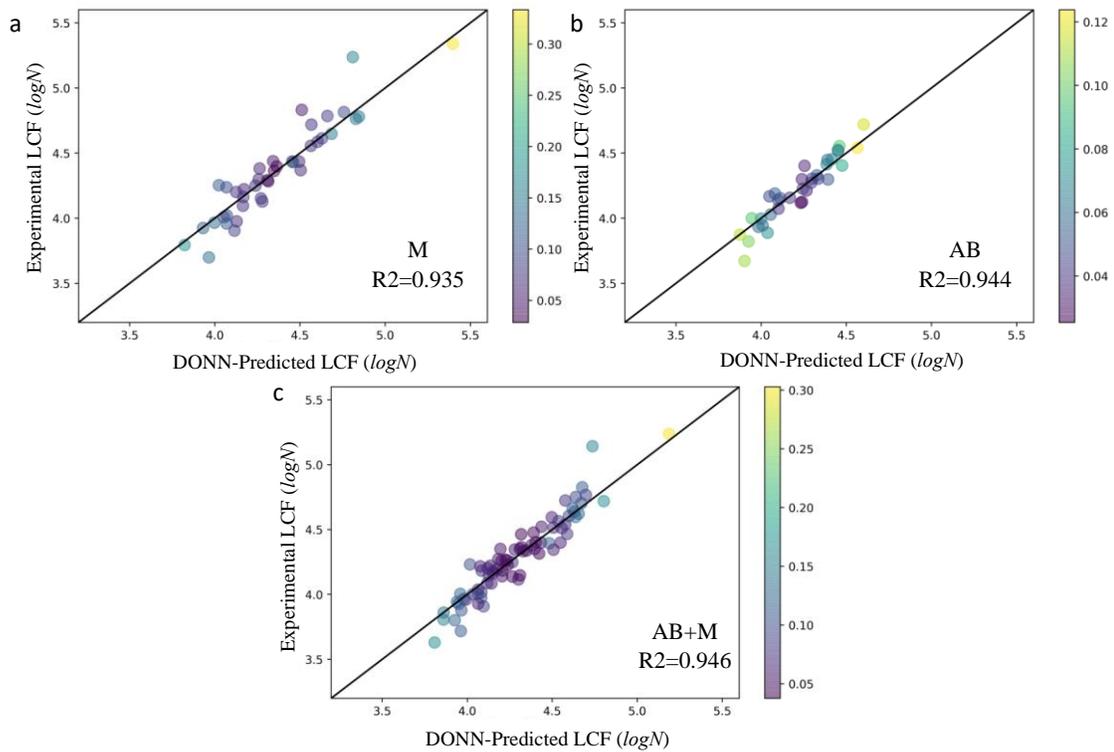

**Figure 16.** Pair plots between DONN predictions and experimental LCF in *logN*, with uncertainties being displayed in different colors for **(a)** AB, **(b)** M and **(c)** combined AB+M samples.

## 4. Conclusion

In this work, we investigated the effects of surface roughness and pore characteristics on the stress-controlled fatigue lives of direct LPBF-printed Ti64 fatigue bars, and developed machine learning models to describe their correlations. The unique feature leveraged in this study, the depletion zone achieved through the contour laser scan, played an essential role in separating the effect of pores and surface roughness. The contour laser scans in the LPBF process make AB samples have similar surface roughness, but diverse internal pore features were achieved by varying the laser scanning speed during the hatch scans. According to the linear-log model, narrow distribution ($R^2 = 0.924$) for all AB samples was presented in the S-N plot. Therefore, this result suggests that the fatigue life of the AB samples is dominated by the microscale surface roughness ($R_a \sim 7.7$ μm) regardless of the internal pore features. The M samples, which have internal pores exposed to the surface after machining, exhibit more scattered S-N plots among samples printed with different laser speeds. This result suggests that the fatigue life of the M samples is largely impacted by the pore features, which are influenced by the laser speed during the LPBF process.

A machine learning model using DONN was established to predict the quantitative relationship between the surface roughness, pore features and the fatigue data. The DONN-predicted average fatigue life agreed well with the experimentally measured values, with Pearson Correlation Coefficients of 0.935 and 0.944, respectively, for the M and AB samples. DONN also has the unique capability of estimating the prediction uncertainty. The estimated prediction uncertainties were below 0.35 for the M samples and below 0.13 for the AB samples. Therefore, we expect that our data-driven surrogate model will contribute to advancing the LPBF process for industrial adoption by providing a fast evaluation of the acceptance of a printed part without the need for time-consuming destructive tests.

**Acknowledgement**

We would like to thank Rolls-Royce for financial support of this work in association with the V4 Institute.

**Author contributions statement**

T. Luo, R. Billo, R. Attardo, C. Tomonto, M. Glavicic and M. Layman conceived the generic concept. R. Attardo, and C. Tomonto performed 3D printing. M. Nordin and Paul Wheelock performed post-processing and CT scans. S. Moon conducted optical surface scan and SEM. S. Moon and T. Luo analyzed all experimental data. R. Ma, R. Billo and T. Luo constructed NN modeling. S. Moon, R. Ma and T. Luo wrote the main part of the manuscript and all authors reviewed it.

**Competing interests**

The authors declare no competing interests.